\documentclass[10pt,journal,compsoc]{IEEEtran}

\ifCLASSOPTIONcompsoc
  \usepackage[nocompress]{cite}
\else
  \usepackage{cite}
\fi
\ifCLASSINFOpdf
   \usepackage[pdftex]{graphicx}
\else
\fi

\usepackage{
	bold-extra,
	eufrak,
	graphicx,
	amsmath,
	tikz,
	diagrams,
	listings,
	placeins,
	tikz-cd,
	hyperref}
\usepackage[ceqn]{mathtools}
\newcommand{\figura}[4][1]{
	\begin{figure}[ht!]
		\begin{center}
			\includegraphics[width=#1\columnwidth]{#2}
			\caption{#4}
			\label{fig:#3}
		\end{center}
\end{figure}}
\newcommand  {\m}{\mathcal{M}}
\renewcommand{\k}{\mathcal{K}}
\newcommand {\M}{\mathfrak{M}}
\newcommand {\K}{\mathfrak{K}}
\newcommand {\I}{\mathcal{I}}
\newcommand {\st}{\quad s.t.\quad}
\newcommand {\rombo}[1]{\begin{tikzpicture}
	\node [decision] {#1};
	\end{tikzpicture}}
\newcommand\Quad[1][1]{\foreach \Quaddy in {1,...,#1}{\quad}\ignorespaces}

\usetikzlibrary{shapes.geometric}
\tikzstyle{decision} = [diamond, draw, text badly centered, inner sep=3pt,aspect=2]

\begin{document}

\title{Polyline defined NC trajectories parametrization. \Large{A compact analysis and solution focused on 3D Printing.}}

\author{Honorio~Salmeron~Valdivieso\thanks{Alexander~Gribov}\thanks{P. Andres-Martinez}}

\markboth{Honorio Salmeron Valdivieso}%
{Shell \MakeLowercase{\textit{et al.}}: Bare Advanced Demo of IEEEtran.cls for IEEE Computer Society Journals}

\IEEEtitleabstractindextext{%
\begin{abstract}
	This paper consists of a formal analysis and one solid solution to the knot finding problem given a source polyline and a parametric curve (e.g. circular arc, ellipse or biarc). We solve the problem using both a greedy algorithm to collect possible arc candidates and a simple algorithm to decide their combination. The rise of 3D printing technology has made it necessary to gain control over how we describe trajectories to our machines. The common method to define paths on 3D printers is describing complex trajectories with high-density polylines. This is computationally expensive and establishes a limit to the greatest accuracy for a given moving speed. This work provides an analysis and a method to fit those polylines with a near-optimal distribution of circular arcs and straight segments.
\end{abstract}

}

\maketitle

\IEEEdisplaynontitleabstractindextext

\IEEEpeerreviewmaketitle

\ifCLASSOPTIONcompsoc
\IEEEraisesectionheading{\section{Introduction}\label{sec:introduction}}
\else
\section{Introduction.}
\label{sec:introduction}
\fi\vspace{3em}

\IEEEPARstart{C}{NC} (Computer Numerical Control) machines are usually controlled using scripts made by consecutive instructions, indicating the machine how to displace the tool, which one to use, etc. The set of displacement related instructions usually includes the capacity of describing some parametric curves as circular arcs. Those are useful capacities because of its low processing impact.\\

However, those instructions are rarely used on CNC 3D printers. It is more common to approximate paths with high segment density polylines (see figure \ref{fig:input}). This is due to the need for automatizing the process of --given an arbitrary 3D design-- to obtain a set of consecutive instructions to print each layer of the shape. Those 3D designs are usually described by meshes of vertices in the 3D space.\\
\figura[0.8]{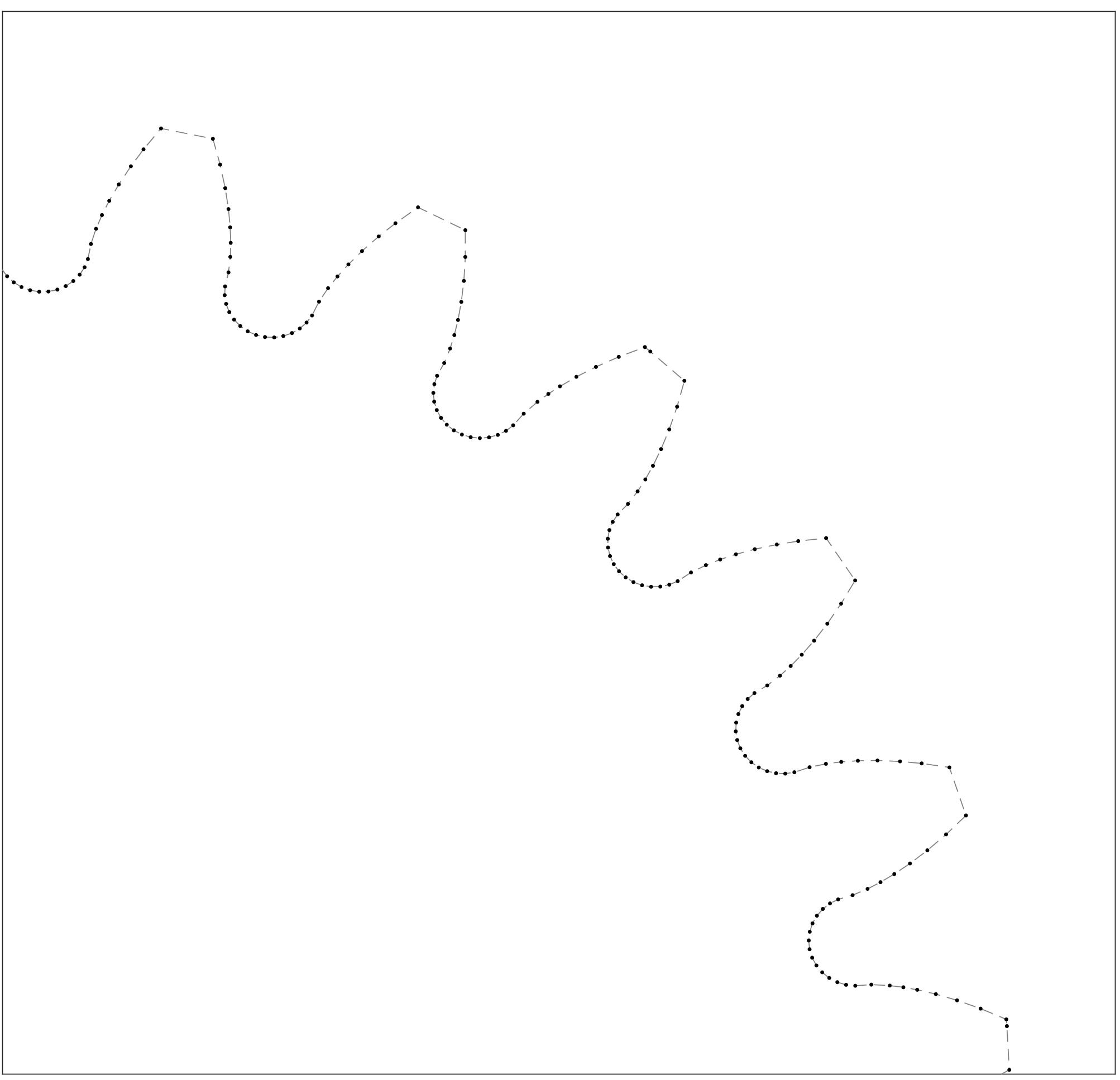}{input}{Detail of a polyline produced by Slic3r for 3D impression. The dots represent the borders of each straight segment, as seen, curves are approximated by many consecutive.}

The main drawback of describing the machining trajectory as consecutive straight segments resides on the limited computational power of most domestic 3D printers for moving data between memories and processing constant flux of information. This establishes a limit over the processed segments per unit of time and, as consequence, a limit on moving speed. It also greatly increases the machine vibrations due to micro-stops on the control flow. This problem is particularly notable on non-cartesian machine geometries (like delta geometry printers) due to the complex cinematic relations between extrusion head and motor movements, which have to be calculated for every instruction.\\

The purpose of this work is to develop a method for, given a polyline defined path, to obtain an approximation made by --both-- straight segments and fitted parametric curves. We also optimize this resultant curve for 3D printing implementation. The problem can be split in two sub-problems: \textbf{knot finding} and \textbf{curve fitting}. First one is related to the optimization of arcs distribution over the solution spline, we use the term \textit{knot} to refer the boundary vertices of those arcs (see figure \ref{fig:knots}). The second sub-problem is the curve fitting itself. This last one has been largely discussed. For example, Milan K.Y and D.J. Walton approached the problem of using biarcs to reach G1 curve interpolation \cite{Milan} but other authors have opted to just use circular arcs and ignore tangent preservation \cite{Ji}\cite{Xi}. The knot finding problem is the most challenging one in the search of an optimal and automated arc distribution over the solution. This topic has been previously discussed for the case of approximation with straight segments \cite{JGD}\cite{Alex4}. Our formalism is independent of both the fitted curve and the fitting method. Different authors have approached the problem of knot optimization with different perspectives. Greedy algorithms like \textit{longest arc}\cite{Alla} and a \textit{bisection method}\cite{Walton} were applied. A dynamic programming approach was also applied in some publications \cite{Ji}\cite{Alex} and evolutionary algorithms have been implemented seeking for optimal distributions \cite{Xi}. 

\figura[0.85]{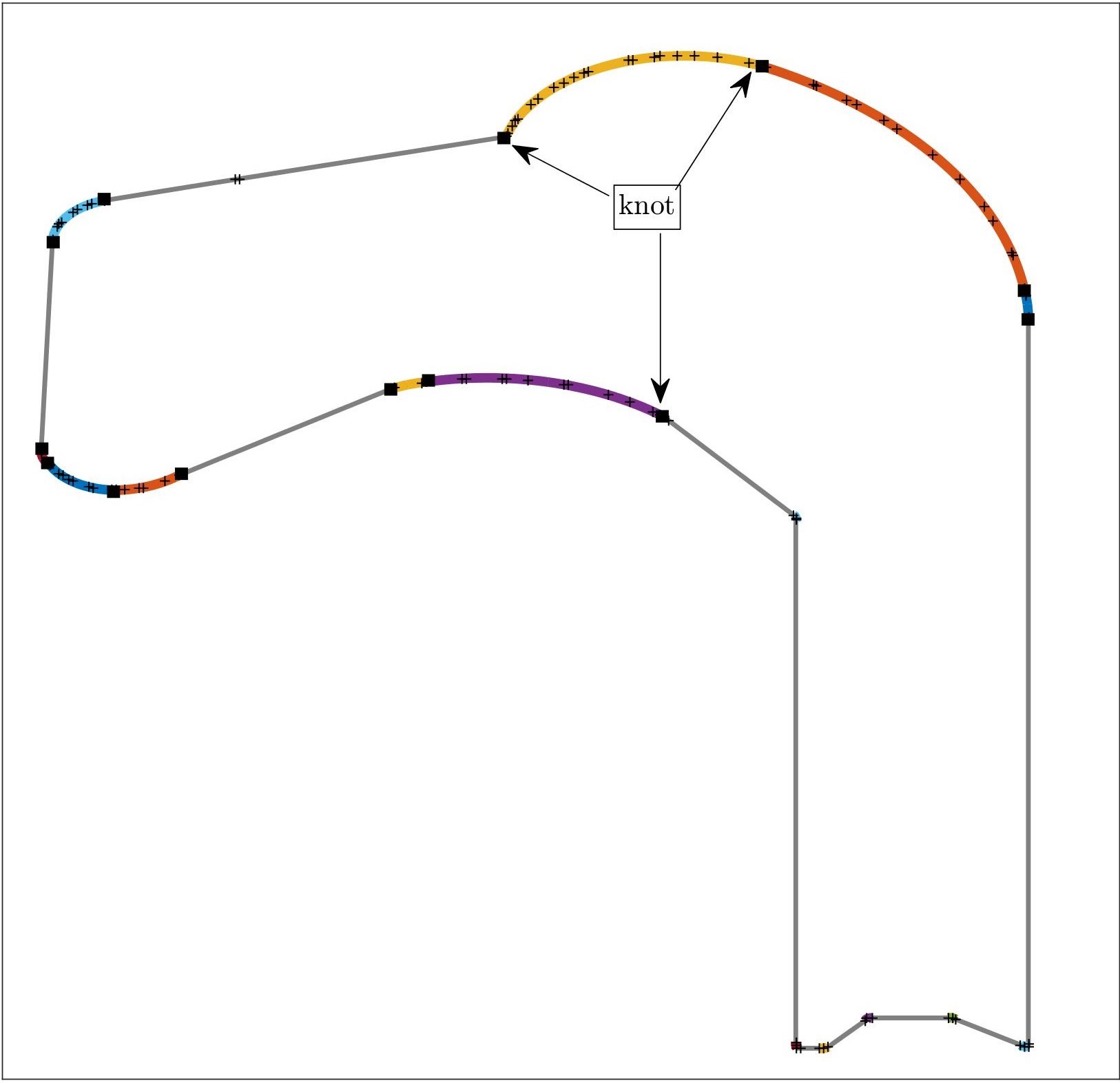}{knots}{The scatter dots (+) represent the source polyline. The coloured parts are the different arcs found.}

Generalized optimal solutions have been published \cite{Alex}. \textbf{Our goal is to obtain an efficient description and solution of the problem for the case of consecutive points and one fitting curve. Our formalism uses sets of source dots to describe each arc domain, allowing a straightforward resolution of the problem and a fine-tuning of the requisites. It also allows us to propose further work on optimization in terms of the description we establish.}\\

\figura[0.85]{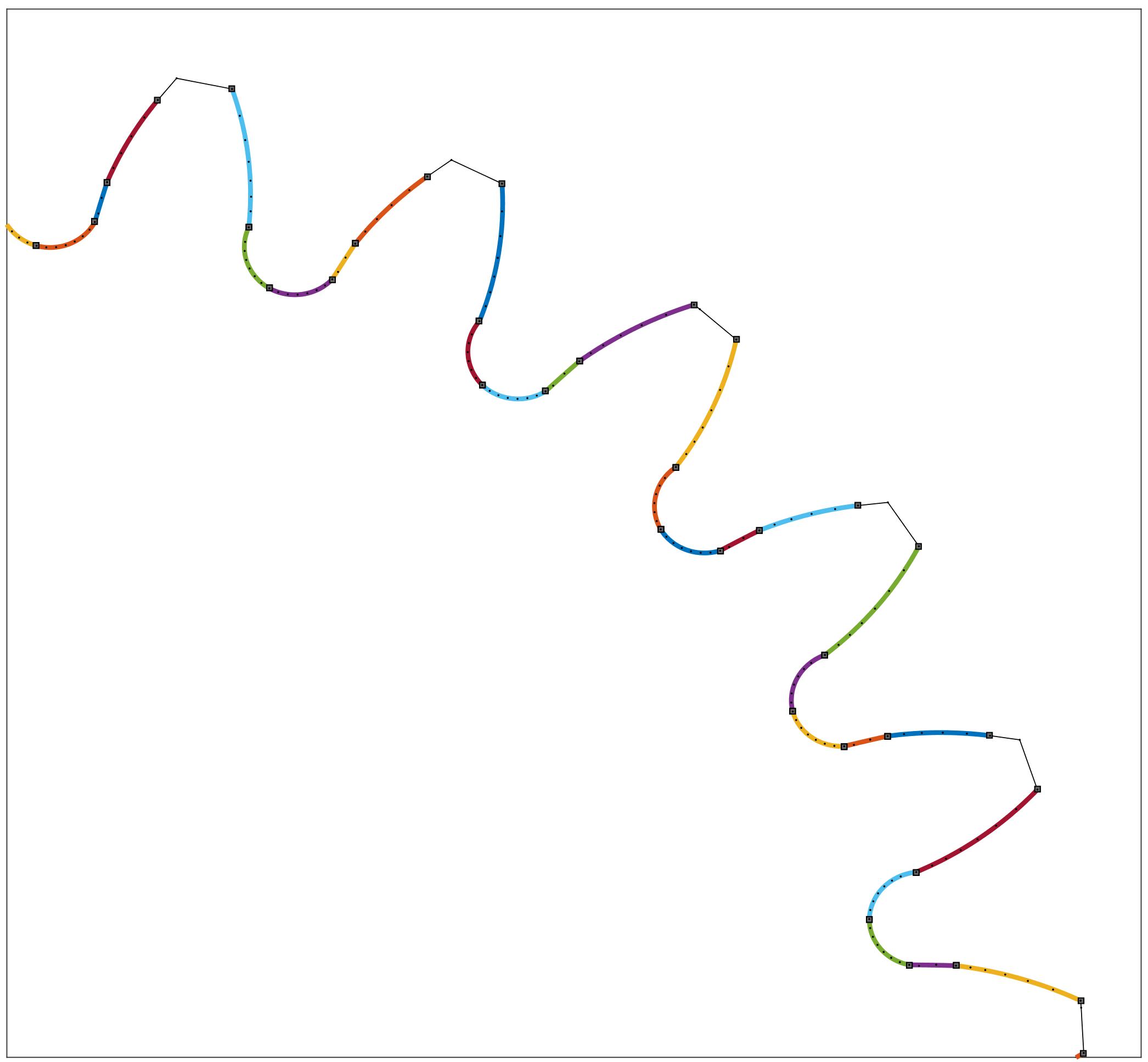}{Result1}{Processed result of input poly-line presented on figure \ref{fig:input}. The colored parts are the different arcs found.}\vfill\eject

In this article we propose: first, a mathematical description of the knot finding problem in the case of using both straight segments and parametric curves. Secondly, a self-designed algorithm to obtain a reasonably optimal solution in a reasonably low computational time. Finally, some technical details and requisites of the final solution (see figure \ref{fig:Result1}) for the case of 3D printers will be discussed. The code designed and applied on this paper result's can be found at \textit{https://www.mathworks.com/matlabcentral/fileexchange/\\68439-polyline-approximation-with-arcs-and-segments}

\section{VERBAL DESCRIPTION.}\vspace{2em}

The input for our resolution algorithm is a set of indexed dots as the one shown in the figure \ref{fig:input} is received. As output, it is expected to obtain a set of (also indexed) straight segments and approximated intervals. It is desirable to avoid collinear dots on the source polyline. It is trivial to remove them and it makes the implementation easier. In our implementation, we make use of circular arcs as approximating curve. Those provide good enough solutions for our final purpose.\\

Our objective is to find, in some input like the named example, the best possible valid distribution of circular arcs over our source polyline (e.g. see figure \ref{fig:Result1}). In figure \ref{fig:input} the beginning of the trajectory and its end does not coincide, but a closed curve can be also computed. We only consider non-closed trajectories for the mathematical analysis since it would be too messy to describe formally.\\

We are focusing on the use of this algorithm for NC machine trajectory optimization, some optimization goals that appeal to this implementation are:\\

\begin{itemize}
\item Maximize the number of source dots approximated by arcs.
\item Minimize the total number of arcs for the same approximation.
\item In order to maximize the time interval between instructions (to avoid CPU overloading), to reduce length difference between consecutive arcs is also desirable.
\end{itemize}

Other control parameters may be added as input for better quality monitoring. Those parameters could be:

\begin{itemize}
\item Maximum distance between two consecutive dots approximated by same arc ($D$).
\item Minimum dot quantity required on any approximation arc ($R$).
\item Maximum collinear error allowed ($\epsilon$).
\end{itemize}

\textbf{The way we propose to search for a solution is: we should first find all the possible valid arcs. Then we must force some conditions in order to keep only the most relevant ones and, in the end, we analyse each intersection between arcs to decide how re-distribute them.} Different steps are going to be described using supersets obeying more and more requisites. The formalism is developed to provide a mathematical structure.\vfill\eject

\section{FORMAL DESCRIPTION OF PROBLEM AND SOLUTION.}\vspace{2em}

While reading the description, assist yourself with figure \ref{fig:example21}.\\ A mathematical description of input (source polyline), output (resulted approximation) and a procedure is developed. The description makes use of fittable intervals of the source. Merging obtained arcs with non approximated parts of the source is post-processing.\\

The main goal of this description is to establish a language to think and communicate effectively the terms of the problem and the requisites we apply in our solution.\\

Call $\Gamma$ to the set of sorted dots introduced as input, then call $\I$ to the indexation set.

\begin{subequations} \label{eq:I}
\begin{align}
&\Gamma=\{\bar{x}_i\}_{i \in \I}\\
&\I={1,2,3...,N}\\
&I_{a,b}=[a,b] \subseteq \I : a\leq b
\end{align} 
\end{subequations}

Where $\bar{x}$ is the coordinate of a dot and $N$ is the total number of dots given as input. we will refer to an arbitrary subset $I_{a,b}$ just as $I$.

Our procedure commences obtaining the superset of all fittable $I$ so we can reduce it to a valid solution. We call this --non restrictive superset-- $\M$.\footnote{Although $I$ does not define spatial location but a location in the input index and would be incorrect to speak about \textit{spatial approximation of $I$}, this is the way we'll referee to the approximation by an arc of $\{\bar{x}\}_I$}.

\begin{subequations}\label{eq:U}
\begin{align}
&\M=\{I \in \I \mid\quad E[I] \le \varepsilon, |I| \geq R,\nonumber\\
&|\bar{x}_n-\bar{x}_{n+1}|\leq D\quad\forall n,(n+1)\in I\}
\end{align}
\end{subequations}\vspace{1em}

Where $E[I]$ is the maximum error committed when fitting on $I$ and $\varepsilon$ is the maximum error allowed. The process of fitting an arc to a set of dots varies. For now, just call $E[I]$ to the maximum collinear error committed when fitting $\{\bar{x_i}\}_{i \in I}$ with a certain method. We will discuss it further.\\

A fundamental property of the fitting is that any subset of a valid interval will also commit a valid fitting error. Taking apart the extra parameters, we could say that it also is in $\M$.

\begin{equation} \label{eq:prop}
\forall I \subset I' : I'\in\M \Rightarrow I\in\M
\end{equation}

Our perspective is to use this property in order to obtain a valid solution. We will do it obtaining $\M$, applying some interesting requisites (the restricted set will be $\m$) and reducing the product to a valid approximation $\k$.

In this formalism, the knot finding algorithm is equal to the process $\M \rightarrow \m \rightarrow \k$. Definitions for $\m,\K$ and $\k$ are presented (in such order), followed by a procedure for obtaining $\m$ and then apply $\m\rightarrow\k$.\\

\begin{subequations}\label{eq:M}
\begin{align}
&\m\subseteq\M \st \forall i\not =j,k:\not\exists \m^i \subseteq (\m^j \cup \m^k)
\end{align}
\end{subequations}\vspace{1em}

Where $\mathcal{M}^i$ represents the ith element of $\m$. The set condition establishes that no element $\m^i$ is completely contained in any other subset $\mathcal{M}^j$ because it wouldn't add useful information. It also discards any $\mathcal{M}^i$ contained in the union of other two ($\m^j \cup \m^k$), because it is never going to be best option to maintain the contained subset than the two subsets containing it. \\

Now we define first, the properties of a generalized solution superset $\K$ and then, the reachable solution superset $\k$. In order to define $\K$ in an efficient way, we need to describe the \textit{interior} of our intervals, this is, the subset without the dots at the borders \footnote{It's understood that a discrete set has no borders, but let us define first and last dots as border.}:

\begin{align}\label{eq:K}
& \star I = I \setminus \ \partial I \st \partial I_{a,b}=\{a,b\}\\
& \K \subseteq\M \st  \forall i\not = j:\star\K^i\cap \K^j= \emptyset
\end{align}\vspace{1em}

The last definitions establish the minimum conditions needed for $\K$ to be a valid solution: To be a set of subsets $I\in\M$ which intersections only happen at the boundaries --only the end and/or beginning of an arc is allowed to be the end/beginning of another--.\\

Have in mind that, since we get the solution from manipulating $\m$ and $\m$ is restricted, we can't achieve any possible solution but only the ones reachable from the information contained in $\m$. For having a correct description of the available solutions, we should add the condition of been reachable by reducing components of $\m$.\\

We call this set $\k$. It has the properties of $\K$ but adds the just named condition.

\begin{equation} \label{eq:k}
\begin{split}
\k\subseteq\M \st \forall i\not = j:\star\k^i\cap \k^j= \emptyset,\\\forall \k^i\exists \m^j : \k^i \subseteq \m^j
\end{split}
\end{equation}

With the last definition, we force $\k$ to be reachable by reducing (or not) every $\m^i$ to a subset of itself.
\vfill\eject

\figura[1]{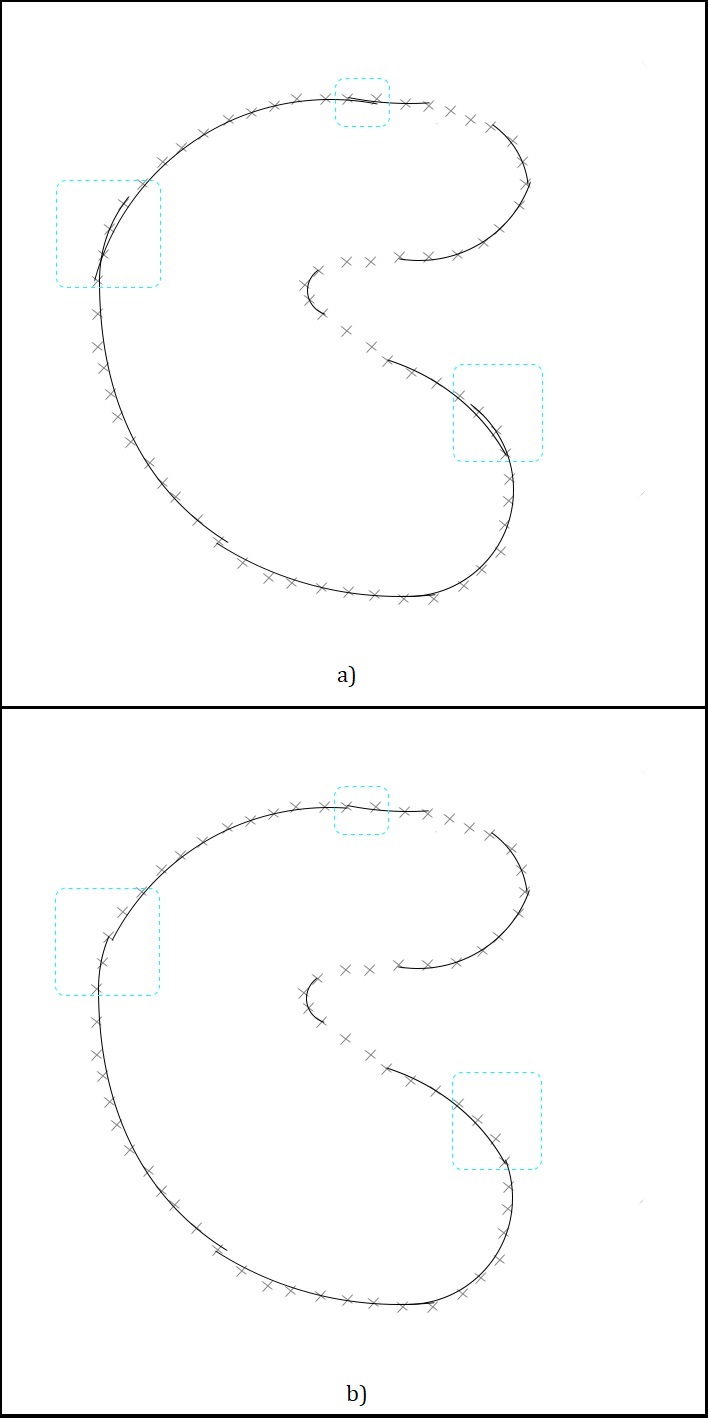}{example21}{\textbf{Figure a):} partial solution described by $\m$. \textbf{Figure b):} final and definitive solution described by $\k$. The intersections are highlighted with squares. Can be seen in \textbf{a)} that there exist intersections which need to be solved with certain criteria in order to obtain in \textbf{b)}. In our case, that criteria are the matching of the optimization points described in section 2.}\vfill\eject

\subsection{Procedure to obtain $\m$, ($\Gamma\rightarrow \m$).}\vspace{2em}

As said, $\m$ contains all the information needed to reach any possible solution $\k$.
A technical implementation of $\Gamma\rightarrow \m$ is quite simple. There is plenty of ways for obtaining $\m$, in this sub-section we expose the method we have implemented, which we find quite efficient.\\

We should run across every dot of $\Gamma$ and find the biggest valid arc starting in that dot, then store it. You should try to do this process while also checking the requisites named on the definition of $\m$. There are two possible implementations: find the arcs and check their usefulness at the same time or do it after obtaining $\M$. We decided to only discriminate --in the code below-- sets contained in other sets and then also discriminate sets contained in the union of other two sets during the intersection finding.

Let's describe the valid sets using its starting points and their extension in the forward direction.  $I_{n,{n+\delta_n}}$.\vspace{1em}

\begin{lstlisting}[mathescape=true,caption={Schematic of the code used on the construction of $\m$.},captionpos=b,frame=single]
function Get $\{\delta_n\}_{n\in\I}$
{ $\delta_1=0$; --> Inizialization.
for n=1 to $|\I|$
{    e=0; --> Inizialization.
while e $< \varepsilon$ and $\delta_n<|\I|$ 
{     $\delta_n++$; --> Add another dot.

if $\delta_n+1 \geq R$;    e=E[$I_{n,\delta_n}$]
--> If the set is big enough to be valid, check collinearity.

if $|\bar{x}_{n+\delta_n}-\bar{x}_{n+\delta_n-1}|>D$; e=$\varepsilon$
--> Check the distance with the just added dot.
}    

if $\delta_n+1 \geq R$; $\delta_n=\delta_n-1$ else $\delta_n=0$
--> If after archiving the boundaries of the interval the size is big enough, keep it removing the just found incompatible dot. 
-->If it doesn't, remove the set.

$\delta_{n+1}=\delta_n-1$;
-->Last line is a optimization tweak: we directly make the algorithm check if the immediate consecutive arc set is large enough to overcome its predecessor. 
-->It allows the code to skip checking all along contained sets.
-->This also means that we only have to initialize $\delta_1=0$.
}
}
\end{lstlisting}\vfill\eject

\subsection{Procedure to obtain $\k$, ($\m \rightarrow \k$).}\vspace{2em}

In order to get $\k$ we have to take care of the intersections between members of $\m$. The intersected parts needed to be solved are the ones between the interior content of the arcs (Arcs are allowed to intersect only at the borders). Those relevant intersections are described in the superset $G$.

\begin{equation}\label{eq:G}
G=\{G_{i,j}=\m^i \cap \m^j\mid\quad \star \m^i\cap\m^j\not = \emptyset  \}_{i\not =j}
\end{equation}

The objective is to modify $\m$ until $G=\{\emptyset\}$. One method could be e.g. completely removing all the internal intersections leaving no approximation there. This method is fast but won't provide an optimal or suboptimal solution.
As said in section 2, we seek some optimization points to be balanced at the same time, let's write them down again in a mathematical manner and by priority order:

\begin{subequations}\label{eq:conditions}
\begin{align}
&\text{Maximize dot covering}\Rightarrow max[|\cup \k|]\\
&\text{Minimize number of arcs}\Rightarrow min[|\k|]\\
&\text{Minimize arc lenght difference}\Rightarrow max[\sum \sqrt{|\k^i|}]
\end{align}
\end{subequations}

Reaching the optimal solution that best fit with those conditions implies solving all $G$ at the same time with an implicit implementation. This happens because every $\m^i$ may be joining zero, one, or even two intersections (one at each side).\\\\

We have leaned towards a simpler solution, in which every intersection is solved independently taking care of the optimization locally. Solving the intersection means finding an agreement dot where the boundaries of the modified sets are going to coincide. It is trivial that this dot can only exist in the interval occupied by the intersection because we are only allowed to reduce each set to a subset of itself, in order to maintain its validity. The next diagram shows the algorithm flux we should apply to every intersection $G^n$.\vspace{1em}

\begin{tikzcd}[column sep=-20pt,cramped]    
G_{i,j}=\m^i\cap\m^j    \arrow[d]        &                                        &\\
\rombo{\text{Posible agreement?}} \arrow[dd,"True"] \arrow[dr,"False"]            &&\\
&\text{Remove smaller set.}        &\\
\text{Find optimal agreement dot}\arrow[d]        &                                &\\    
\text{Modify $\m^i$ and $\m^j$}\\[-17pt]\text{to make them match}\\[-17pt]\text{only at agreement dot.}&&
\end{tikzcd}\vspace{2em}

The first part of the algorithm consists in reasoning if do exist some interval inside the intersection where agreement is possible (possible means that resultant arcs are valid). If an agreement is possible, then we should decide, in such interval, which dot is the best for our optimization requisites.
Every part of that diagram is going to be both described and solved. We need the following definitions to describe our particular way of solving every local intersection. We basically need to describe the size reduction on both members of the intersections ($S^i,S^j$) and the maximum possible reduction for each member ($S^i_{max},S^j_{max}$).

\begin{subequations}\label{eq:def}
\begin{align}
& \forall G_{i,j}, |\m^i|\geq|\m^j| \\
& \nonumber \\
& S^i_{max}=max \quad |I| \st I\subseteq (\m^i \cap \star G_{i,j}),\m^i \setminus I\in\M\nonumber\\
&S^j_{max}=max \quad |I| \st I\subseteq (\m^j\cap \star G_{i,j}),  \m^j \setminus I\in\M\\
&\forall I,I' \st \m^{i*}=\m^i\setminus I, \quad \m^{j*}=\m^j\setminus I',\nonumber\\
&\Quad[6]|\m^{i*}\cap\m^{j*}|=1\nonumber\\
&\Quad[6]\Rightarrow  S^i=|I|, S^j=|I'|=|G_{i,j}|-S^i-1
\end{align}
\end{subequations}

$\m^{i*}$ represents the modified ith set. \ref{eq:def}a establishes that, for an intersection $G_{i,j}$, we'll call $\m_i$ to the biggest set implied. \ref{eq:def}b establishes the maximum value that $S^i$ or $S^j$ could take. Calculate those is trivial, we just have to ask both solution sets to join the intersection while having their minimum valid size.\\

\ref{eq:def}c actually means \textit{if agreement is possible, and we call $S^i$ to the quantity of dots we subtract from the biggest set, the quantity of dots we subtract from the smaller set is $S^j=|G_{i,j}|-S^i-1$}.\\

\begin{subequations}\label{eq:Smax}
\begin{flalign}
&S^i_{max}  =  \left \{
\begin{matrix*}[l]
|G_{i,j}|-1     &if\quad |\m^i|-|G|+1 \geq R\\
|\m^i|-R     &else
\end{matrix*} \right . 
\end{flalign}
\end{subequations}\vspace{1em}

First, we should study the possibility of finding a solution to this re-accommodation. It may be impossible to reduce enough the sets without invalidating (figure \ref{fig:example3}). \textbf{If one dot of $G_{i,j}$ is going to be the accordance point between the resultant sets, the sum of this dot plus the removed dots should equal $|G_{i,j}|$.}\\

Last statement implies that the sum of $S^j_{max}$, $S^i_{max}$ and the agreement dot should be greater than the size of the intersection. The condition for existence of agreement is:

\begin{equation}\label{existence}
|G_{i,j}|\leq S^i_{max}+S^j_{max}+1
\end{equation}

In the case of not existing possible agreement, our decision is to discard the tiniest set.
If we have at least one agreement dot, we should decide which one to take. The agreement that we chase is the one that best fits with our optimization goals \ref{eq:conditions}.\\

The condition \ref{eq:conditions}c is going to be included searching for the agreement that minimizes $|\m^{i*}|-|\m^{j*}|=f(S)$:

\begin{align}\label{eq:size}
|\m^{i*}|&=|\m^i|-S^i \nonumber\\
|\m^{j*}|&=|\m^j|-S^j \nonumber\\
f(S^i)&=(|\m^i|-S^i )-(|\m^i|-S^j) \nonumber\\
&=|\m^i|-|\m^j|+|G_{i,j}|-2S^i-1
\end{align}

Function $f(S^i)$ is bounded by $S^i\in [S^i_{min},S^i_{max}]$. We should check for a minimum in that interval. The minimum can be at the boundaries or not. Since the function is crescent, we just have to check if it is positive at $S^i_{min}$.

In conclusion for the optimization of the agreement, you should first decide if the minimum is at the boundaries and calculate it:

\begin{subequations}\label{eq:S}
\begin{flalign}
&S^i  =  \left \{
\begin{matrix*}[l]
|G|-S^j_{max}+1&\scriptstyle{if\quad |\m^i|-|\m^j|\geq|G|-2S^j_{max}-1}\\
\\
\frac{|G|+|\m^i|-|\m^j|-1}{2} &else
\end{matrix*} \right.
\end{flalign}
\end{subequations}

\figura[0.9]{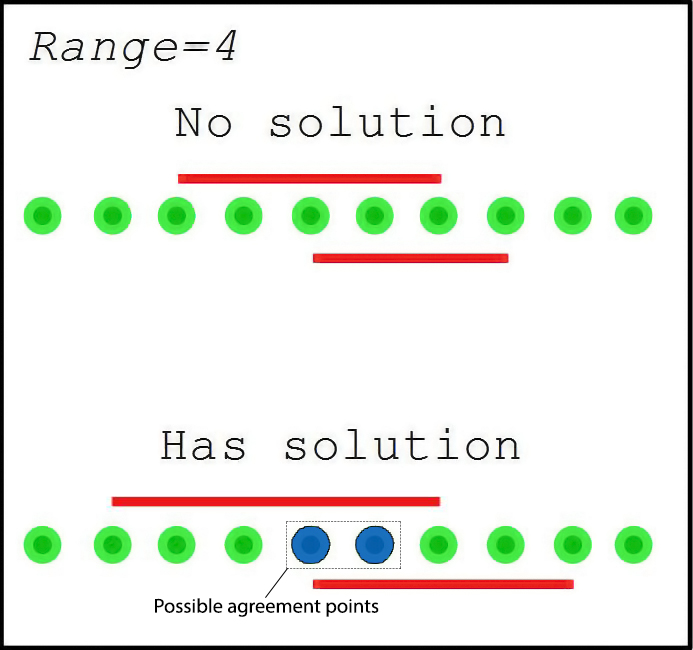}{example3}{Schematic of intersection of dots intervals ($\m^i,\m^j$). The red lines represent both domains. In the first situation there is no reachable solution since there is no valid subset combination of themselves that intersect only at the boundary keeping them both valid sets.}

Having solved the details of the algorithm this is the final flux diagram, which has to be applied for every intersection $G_{i,j}$\footnote{As remember note: i and j are always representing arbitrary non-equal numbers.}. Note that the order in which the intersections are solved affects the final solution.\vfill\eject

\begin{tikzcd}[row sep=scriptsize,column sep=tiny,cramped]
G_{i,j}=\m^i\cap\m^j\arrow[dd,"\text{Evaluate possible agreement.}"]&\Quad[3]                                  &\\
&                                                            &\\
\rombo{$|G_{i,j}|\leq S_{max}^j+S_{max}^i+1$}\arrow[dd,"\text{True, Minimize f(S).}"]\arrow[dr,"False"]&    &\\
&\m^*=\m \setminus \{\m^j\}                                    &\\
\rombo{\scriptsize{$|\m^i|-|\m^j|\geq|G_{i,j}|-2S^j_{max}-1$}}\arrow[dd,"True"]\arrow[ddr,"False"]&        &\\
&                                                            &\\
S^i=|G_{i,j}|-S^i_{max}+1\arrow[dd]        &S^i=\frac{|G_{i,j}|+|\m^i|-|\m^j|-1}{2}\arrow[ddl]        &\\
&                                                            &\\
\m^{i*}=\m^i\setminus I\\[-15pt]\m^{j*}=\m^{j*}\setminus I'
\end{tikzcd}\vspace{2em}

\subsection{Searching for intersections.}\vspace{2em}

The process of looking for intersections may become an important computational load. The way we do it is to evaluate all $\m$ from the biggest to the smaller and look for intersecting sets on their right. \vspace{2em}

\begin{tikzcd}[row sep=scriptsize,column sep=-55pt,cramped]
\text{Look for the}\\[-15pt]\text{biggest $\m^i$}\arrow[d]    &\Quad[15]     \\
\text{Look for intersections on its \textbf{\textit{right}}}\arrow[d]    &        \\
\rombo{Exist intersection?}    \arrow[ddd,"True"]\arrow[ddr,"False"]        &        \\
&        \\
&\text{Discard $\m^i$ from search}    \\
\text{Apply solving algorithm}\arrow[d]                                    &        \\
\text{Discard $\m^i$ from search}
\end{tikzcd}\vspace{2em}

Having obtained $\m$ is equivalent to having obtained $\{\delta_n\}_{n\in\I}$ described in section 3.1. This second concept is closer to the technical implementation and that's what will be used on the next commented schematic.

\begin{lstlisting}[mathescape=true,caption={Schematic of the code used for finding the intersections.},captionpos=b,frame=single]
do
{

$\m^i=I_{n,n+\delta_n}$=FIND_BIGGER($\{\delta_n\}_{n\in\I\setminus Solved}$)
--> We first search the biggest valid set of $\m$ discarding the already processed ones.

$\m^j=I_{n',n'+\delta_{n'}} \st n'\in\m^i: (n'+\delta_{n'})$  is  max
--> Now we have to find the valid set which starts in $\m^i$ and reaches a further dot in the forward direction.
-->Set to zero the rest of intersecting sets.

SOLVE_INTERSECTION($\m^i,\m^j$)

} while $|\m^i|\geq R$
\end{lstlisting}

The reason we only look for intersections on the \textbf{\textit{forward}} direction on the evaluated $\m^i$ is because it represents a low computational load to just check if some next arc starts from the interior of $\m^i$.

After obtaining our solution superset $\k$, the final trajectory can be obtained substituting on $\Gamma$ the intervals of dots included in $\k$ by the correspondent arc.

\section{DISCUSSION.}

A mathematical description of the problem and a proposed solution has been performed. The description uses the indexation of the source dot to describe the domain of each arc. Although this restriction may affect the solution quality, it does the job with good results and much lower computational impact than working on the continuum. The reader should remember that typically, the inputs are high segment density polyline defined trajectories, so it makes no great difference to work using those discrete possible positions than working in a 2D continuum.

\subsection{Post processing.}\vspace{2em}

Post-processing of the arcs parameters can be attempted knowing that a valid collinear error is going to be reachable. Some aspects of the final solution can be taken in care here as, for example, ensuring continuity on the resultant curve or softening the accordance points between arcs.

\figura[0.8]{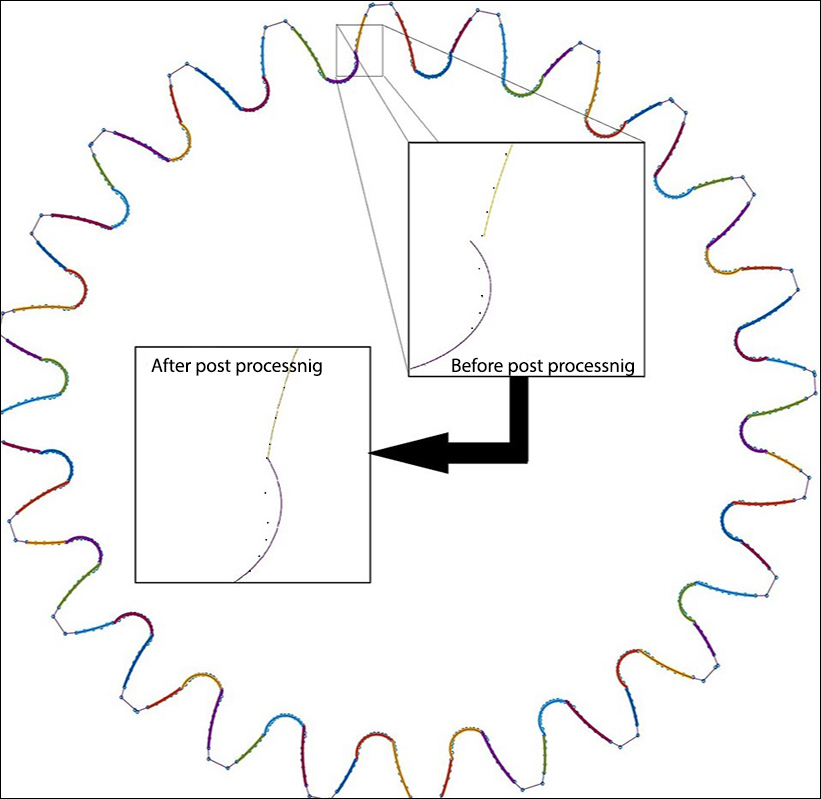}{Post1}{Product of processing a gear-shaped polyline input. Detail of two consecutive arcs before post-processing and after post-processing. A big $\varepsilon$ was used to exaggerate the effects on the accordance point.}

Both aspects have been solved in different ways and exigence levels, some publications as \cite{Walton} show different orders of interpolation for softening the accordances. We decided to just force the boundaries of consecutive arcs to match in a position given by the accordance point. A circular fitting method given ending and starting points is developed by Alexander Gribov \cite{Alex2}.

As a fast solution, three-point fitting (using the middle point and the two at the boundaries) can provide good results if the collinear requisite is high enough.
\subsection{Arc fitting. Error measurement.}\vspace{2em}

In this attempt, we decide to use only circular arcs due to simplicity. The same process with no modification may work for a generalized ellipse, parabola, biarcs...etc. The fitting method we used is the Taubin's \cite{Taubin}. Another one like Kesa's \cite{Kasa} could be faster but implies more memory usage due to the matrix allocation. This makes the process slower because of the time it takes to administrate memory.\\

\figura[0.7]{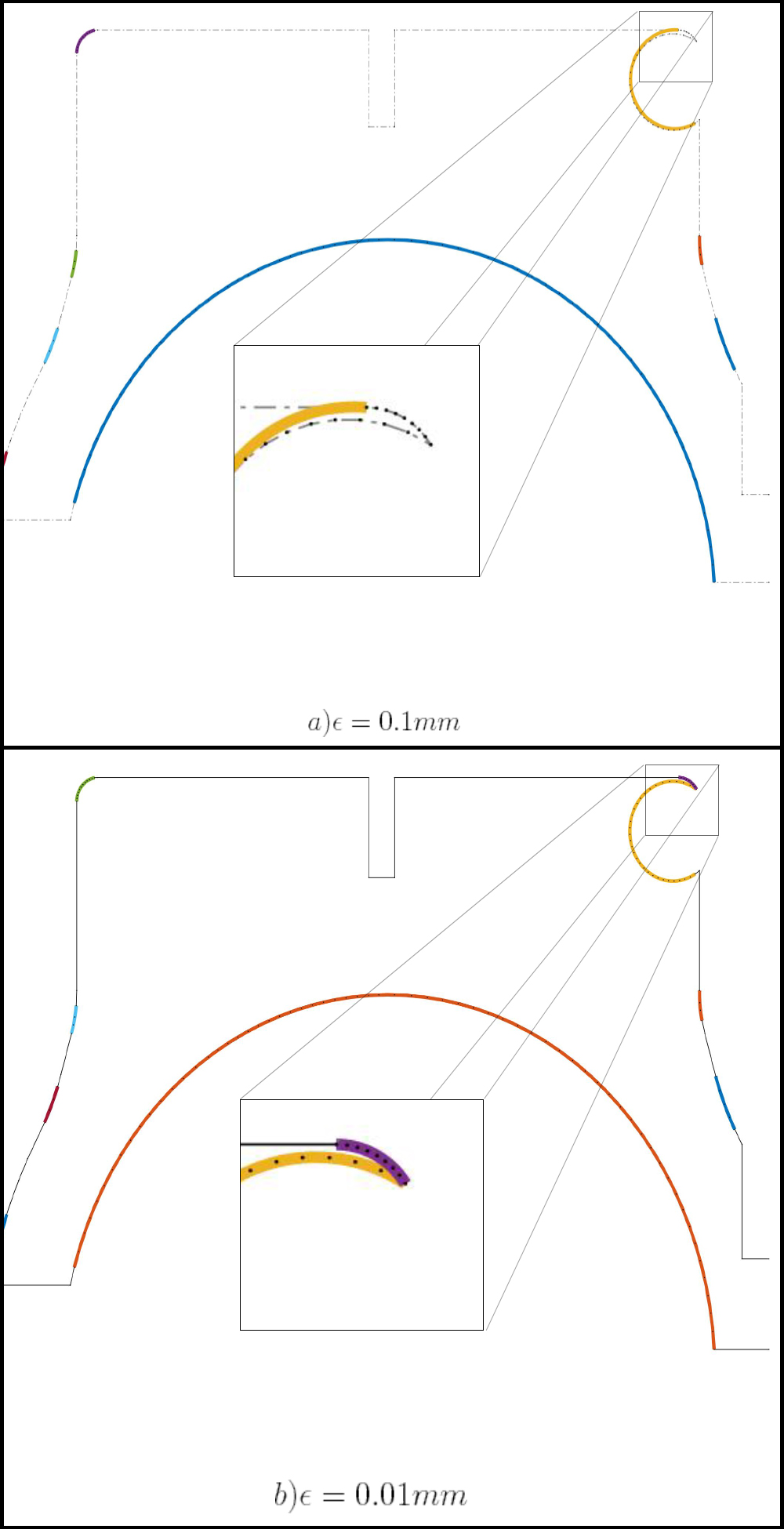}{shapefactor}{\textbf{a)} The big collinearity allowed makes the algorithm to ignore the source shape since the error required is reached. \textbf{b)} Smaller error restriction forces the resultant curve to fit the source shape.}

The way of measuring the error committed by an approximation with an arc on a set of points is discussed in every paper on this topic. Different methods have varied implications. We used as $E[I]$ the maximum absolute difference between the arc radius and the distance from dot to centre. We employed this method due to simplicity since the work isn't focused on the calculation but the knot optimization of the resultant curve. A very important side effect can be distinguished in figure \ref{fig:shapefactor} and is the lack of any measurement of the shape fidelity.

This is solved in the work of Eugene Bodansky and Alexander Gribov \cite{Alex3}, in which the monotony continuity of the source polyline is analysed by the algorithm.

\subsection{Skewness effect of proposed intersection resolution model.}\vspace{2em}

The procedure of finding and solving intersections only at the right of a certain $\m^i$ starting from the biggest one implies some skewness. However, this tendency disappears when increasing the accuracy exigence, because the exact shape of the input becomes mandatory on the final solution shape.

\figura[0.9]{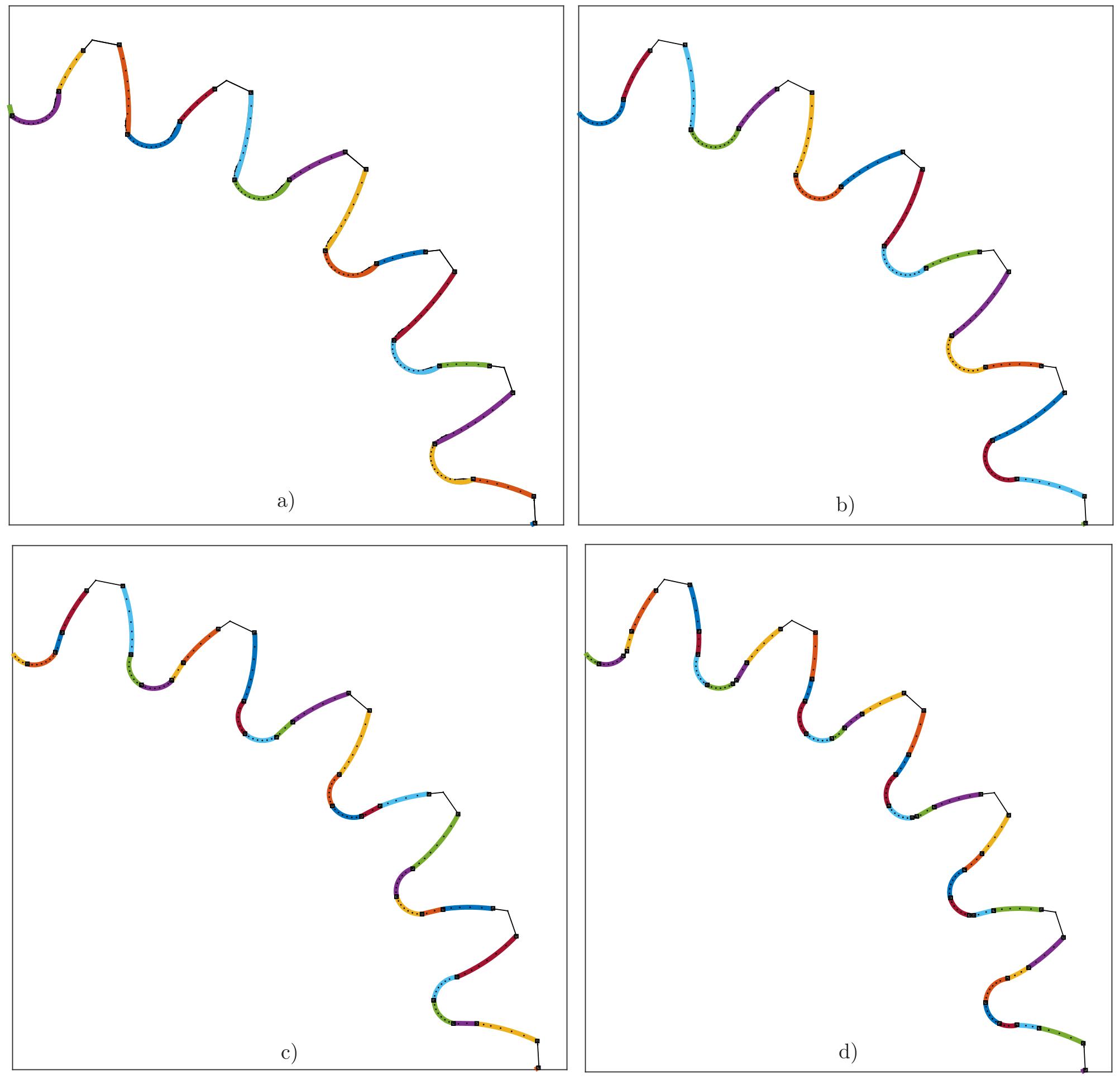}{skewness}{Product of processing a gear shaped polyline input. \textbf{Figure a:} $0.1mm$ collinear error. \textbf{Figure b:} $0.05mm$ collinear error. \textbf{Figure c:} $0.01mm$ collinear error. \textbf{Figure d:} $5\mu m$ collinear error.}

\subsection{Open curves vs closed curves.}\vspace{1em}

The set-based description offered in the last section is only fully consistent if our input is an open curve. The definition \ref{eq:G}, for example, isn't valid if the input is a closed curve since it may happen that $\m^i\cap\m^j$ produce more than one simply connected interval $I$ and $\m^i\cap\m^i\not = \emptyset$ for the case of a circumference input with circumference arcs fitting.\\

However, at the technical implementation, this does not represent a problem since it's easy to generalise the algorithm for open and closed curves. This generalization is a great problem in the process of describing it in a formal set definition fashion but at the technical level can be solved with some tweaks like making your addition and subtraction operations travel between the last and first dots.
\subsection{Continuity between consecutive layers.}\vspace{1em}

If so much collinear error is allowed, it may affect to the continuity on the surface in the direction of the layer deposition. This emerges from the uncertain approximation that affects to every processed layer. This possible \textit{roughness} on the deposition direction can be measured with the allowed collinear error. It has more effect than a simple roughness. It may affect to the adhesion of each new layer if the collinear error is bigger than the deposited filament, because parts of the new layer may fall where no material exists. A value between 0.05mm and 0.01mm should keep good approximations.

\figura[1]{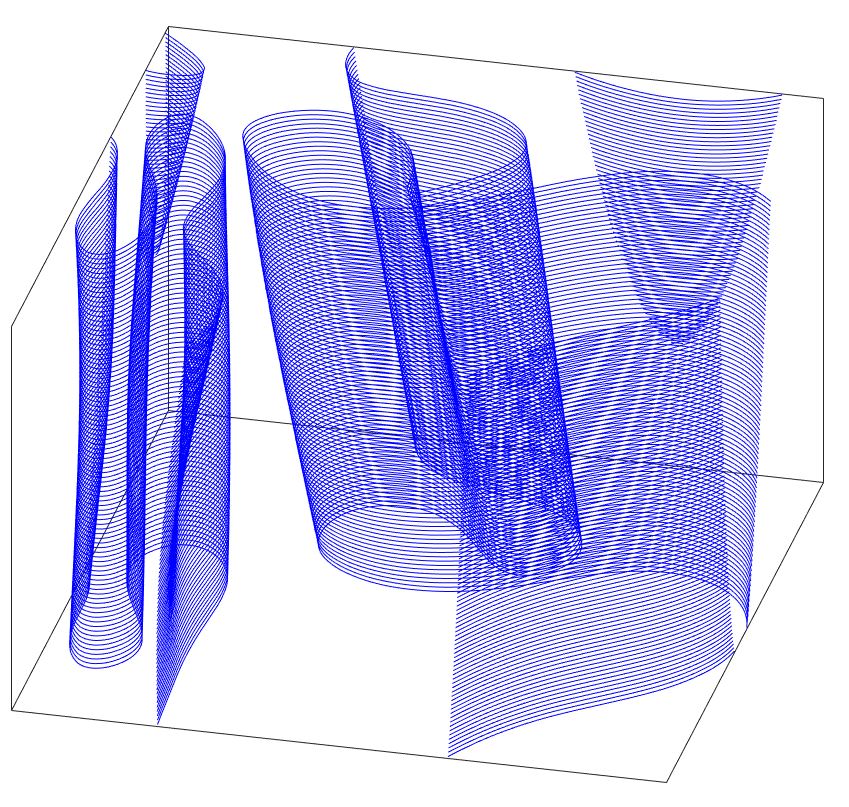}{Simulacion}{Volumetric section of a simulation performed with MATLAB, all the layers of a vase figure were processed and piled together. The collinearity requisite was 0.05 mm, this was the minimum value which ensured soft evolution in the Z direction.}\vfill\eject
\subsection{Comparison with other methods.}

There exist faster methods and higher quality solution methods.
Greedy methods \cite{Alla}\cite{Walton} produce faster solutions but with the cost of big losses on the available information to decide the final solution. Other methods like dynamic programming methods \cite{Alex}\cite{Liu} may find the optimal solution but those are harder to implement and take more time to reach a solution since they don't sacrifice information.

While the dynamic programming approach keeps all the information and the longest arc approach looks for the immediate valid solution, our algorithm stays in the middle using a greedy algorithm to collect arc candidates and them applying simple optimization to decide their boundaries.

\figura[0.7]{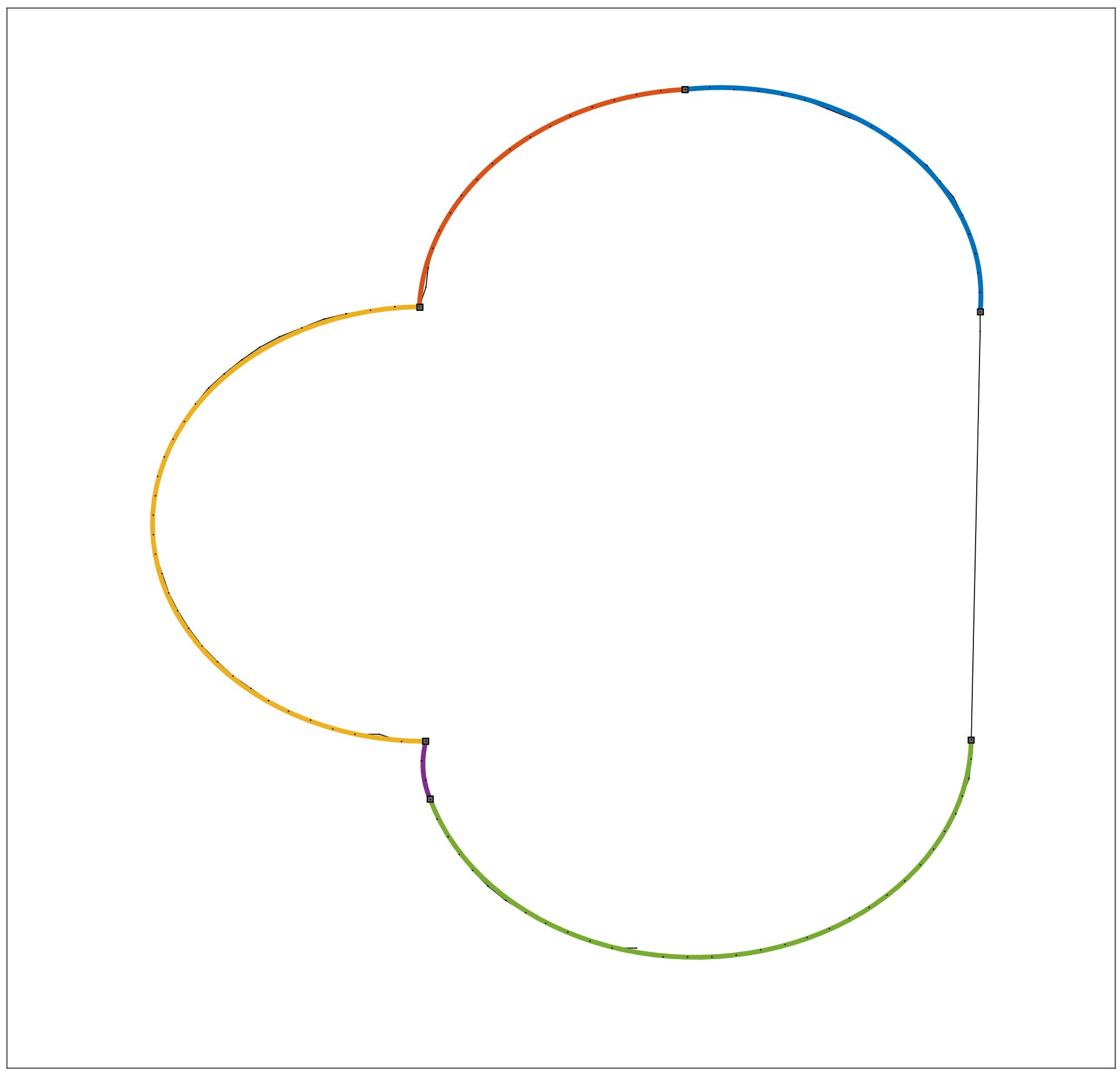}{comp}{Figure reproduced from sample of \cite{Xi} for performance comparison. Solved with a collinearity equal to the 15\% of the source inter-dot distance. They used an evolutionary process with a result of 26 arcs and 1.6 seconds of processing time. We reach the same accuracy with 5 arcs and 27$\mu$s of processing time.}

\figura[1]{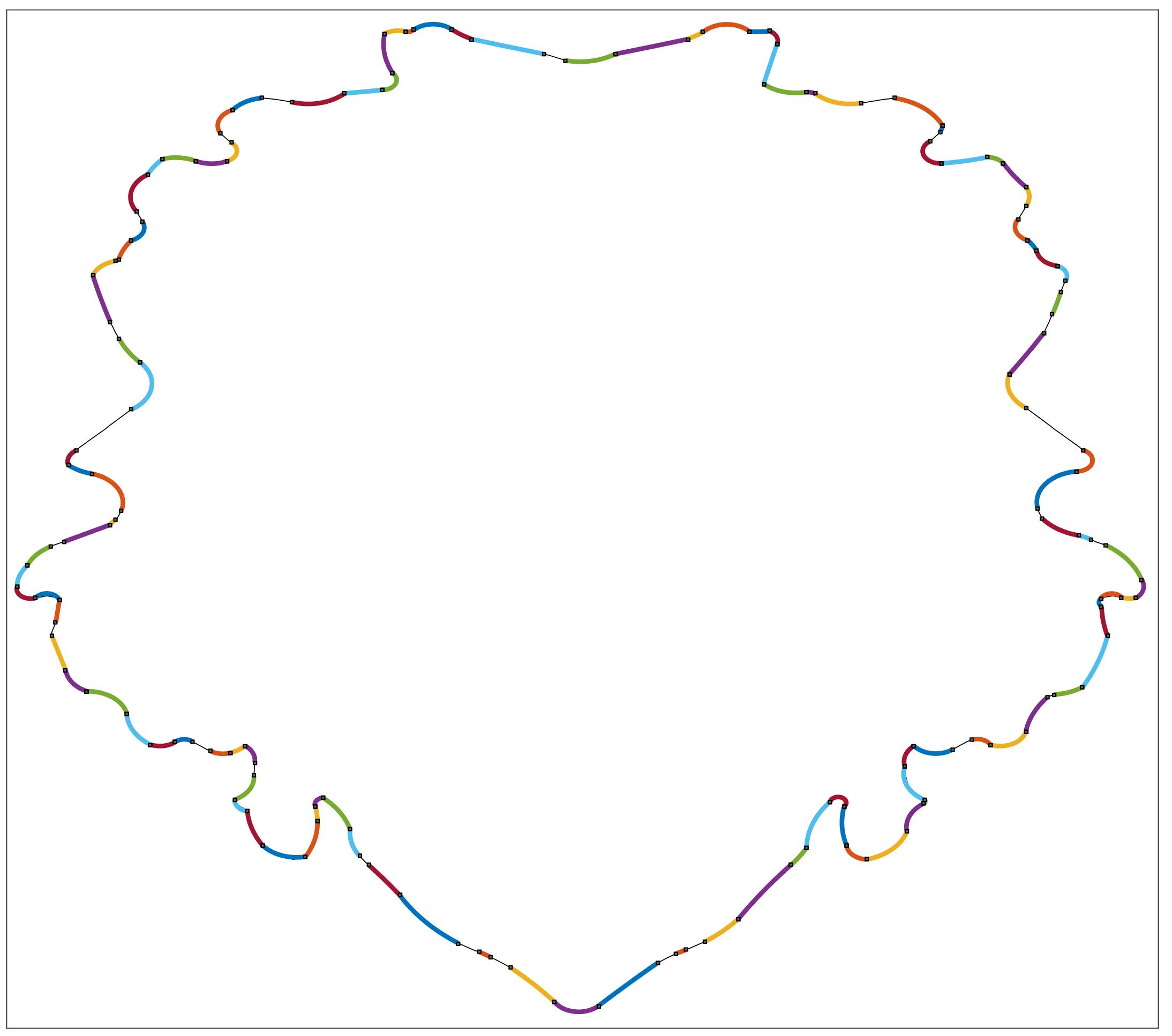}{Sample5}{Sample processed. 109 arcs found, 5ms of processing time. The complex shape of this curve makes mandatory to use $R=4$ to avoid abrupt arcs.} \vfill\eject

\section{RESULTS AND PERFORMANCE.}

High efficiency is mandatory if we want to integrate the method on an automated process for obtaining 3D printer instructions from a 3D design. The performance results obtained by us manifest, by far, good enough time processing marks for high accuracy solutions. The test was performed with a script running in MATLAB installed on an Asus G56JK laptop. No complex tweaks or heuristics were used. Further implementation of lower level languages may increase performance.

\begin{figure}[ht]
\centering
\includegraphics[width=(0.9\linewidth)]{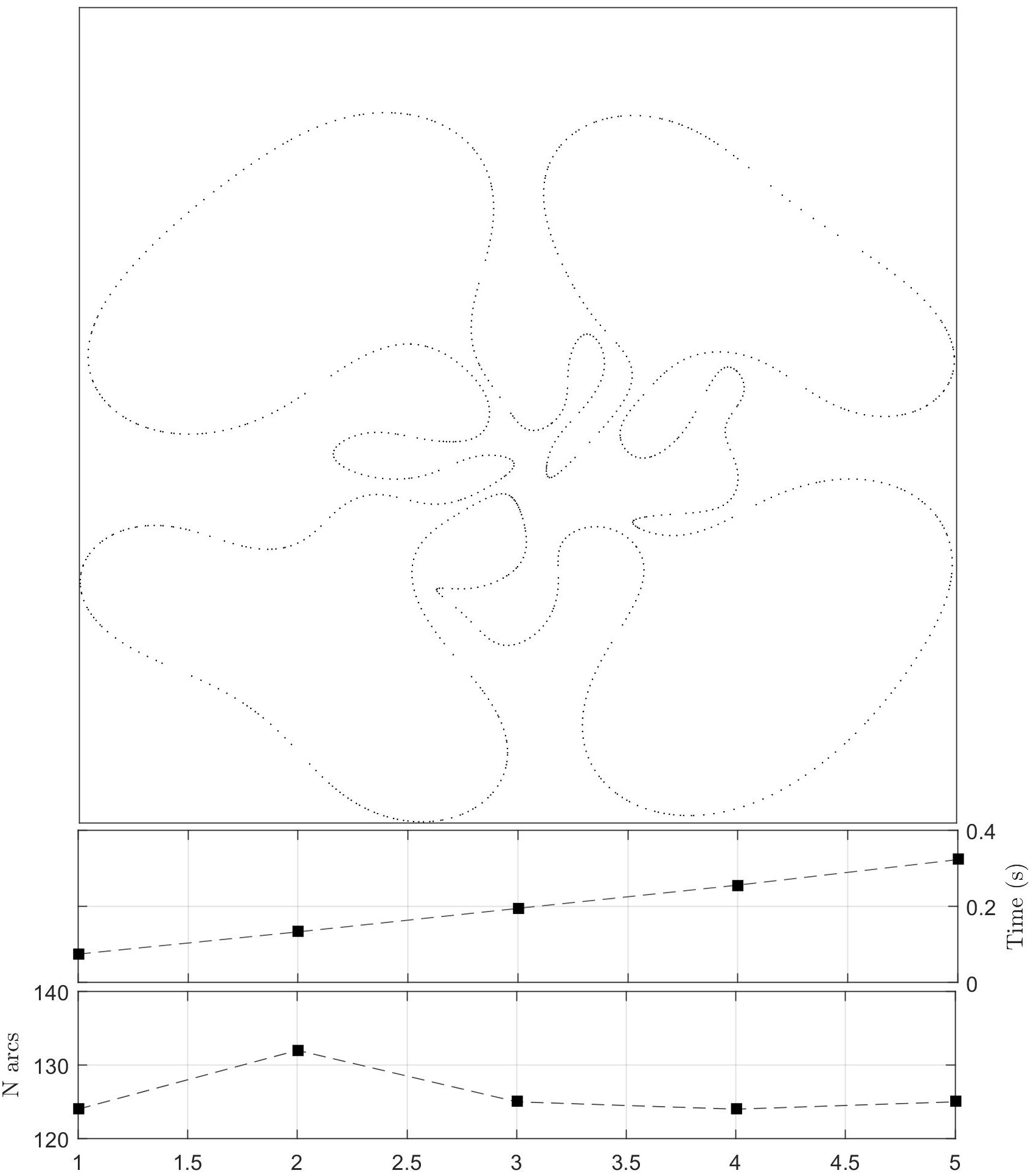}\vspace{1em}
\includegraphics[width=(0.9\linewidth)]{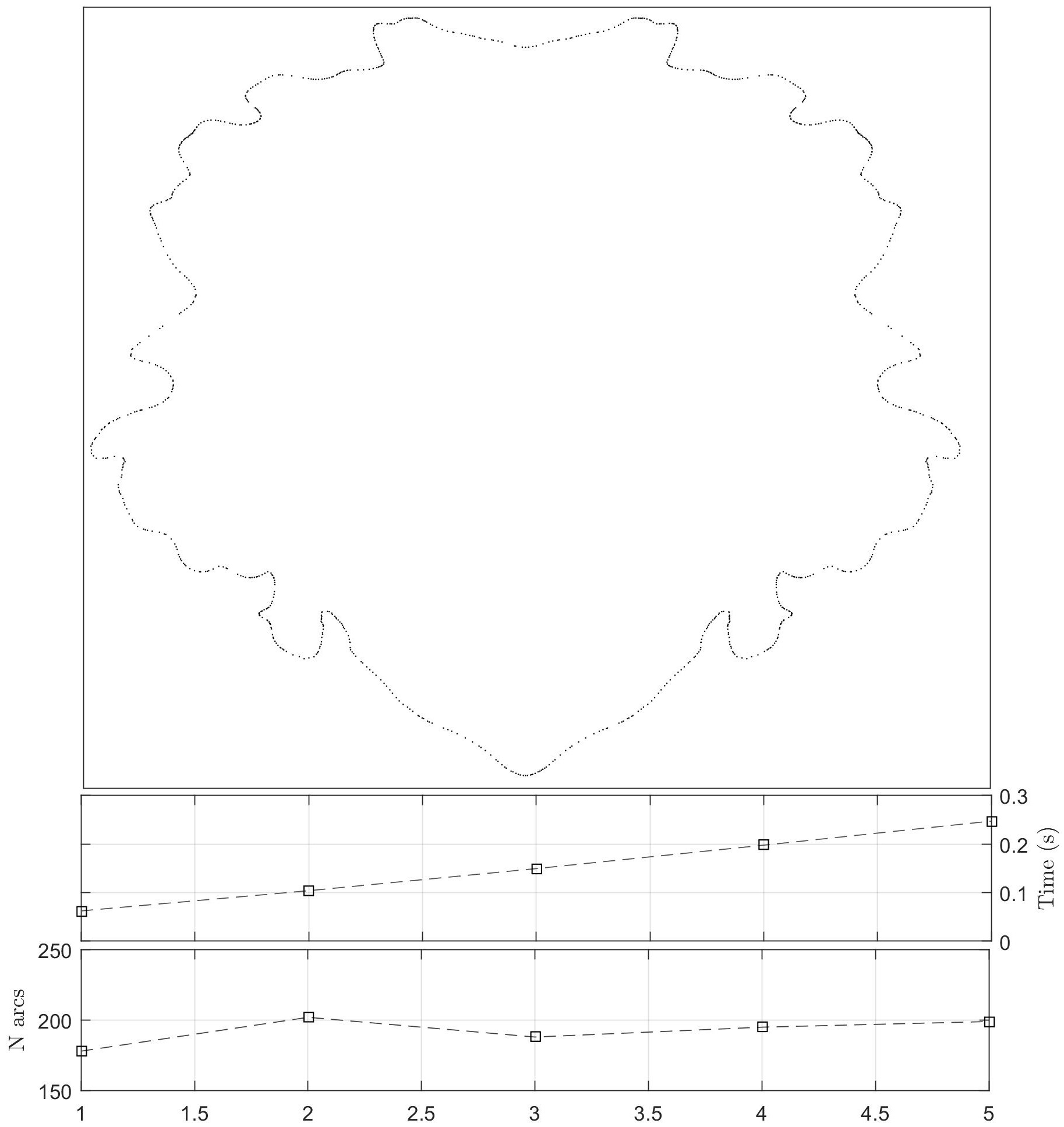}
\caption{Processing time and number of resultant arcs as function of a interpolation factor applied on the source polyline. Processing time grows linear with the number of input dots for same source shape. Number of arcs remains mostly constant as it is expected.}
\label{fig:complexity}
\end{figure}

\begin{figure*}
\includegraphics[width=\textwidth]{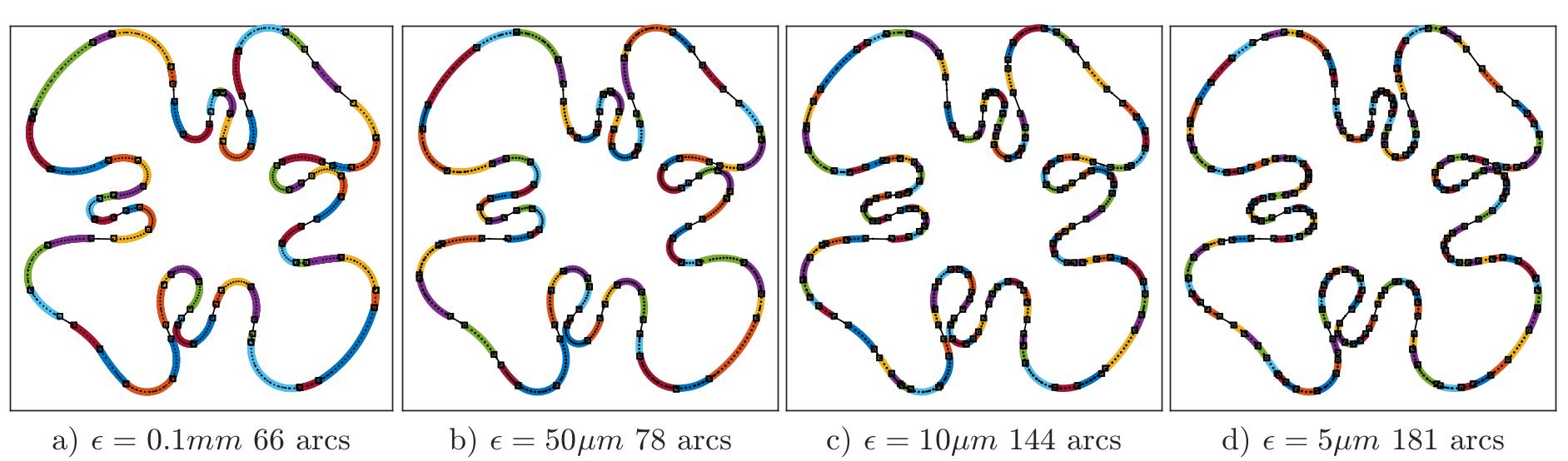}
\caption{A sample solved with different errors. Different arcs are highlighted with different colours (web version).}
\label{fig:perf4}
\end{figure*}
\begin{figure*}
\includegraphics[width=\textwidth]{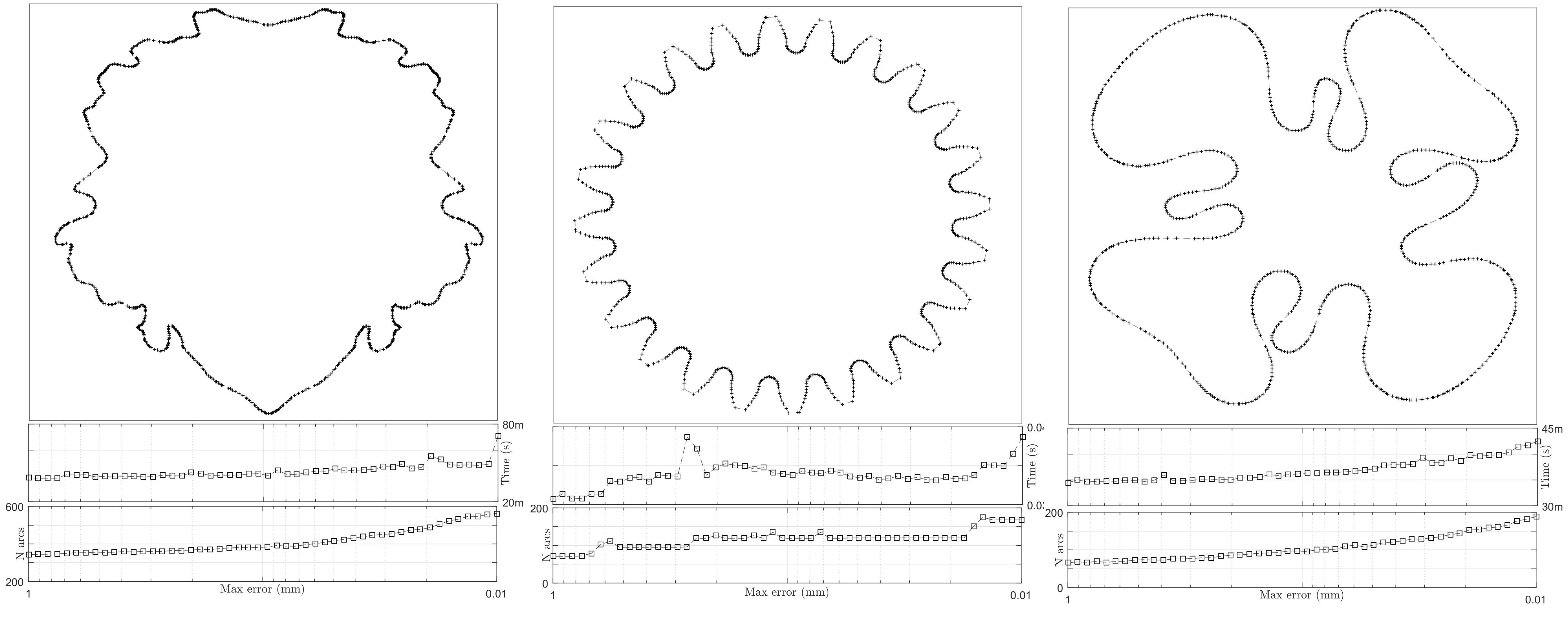}
\caption{Performance test over the shown figures. Processing time and the number of arcs on the final solution versus the collinear error allowed. A minimum range of 4 dots arc was imposed and a maximum inter-dot distance of 0.7 mm. On soft, organic curves (figure c), processing time and the number of solution arcs grow clearly in an exponential manner after a certain value. This is a common situation on 3D printed parts since they are usually designed with parametric CAD design software. On complex, spiky curves (figures a and b), the number of arcs grows mostly exponentially but with fixed steps and the processing time evolves in a less predictable fashion but mostly also exponentially.}
\label{fig:perf1}
\end{figure*}

\section{Conclusion and further work.}

An efficient, simple and compact method for spline fitting has been developed. It is resistant to noisy inputs and always returns a solution.
Keeping smaller tolerances the method always produce good quality results in shape, then it is implementable to a user-friendly software.\\

Although the method is fast and currently useful if the correct control is applied on the input parameters, it does sometimes fail in keeping the original polyline shape because of the lack of some shape geometry penalty on the error function. Developing an implementation of this measurement is intended for further analysis. A good approach to this idea is described in \cite{Alex3}.

Also, a dynamic collinearity detection of groups of dots can be performed analysing the parameters of the resultant arcs.

\vfill\eject
\vfill\eject

%
\begin{IEEEbiography}[{\includegraphics[width=1in,clip,keepaspectratio]{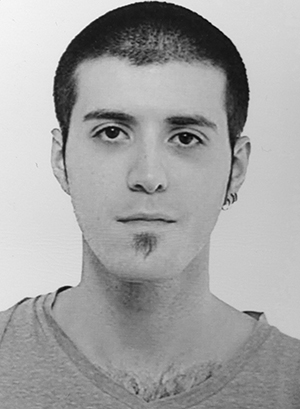}}]{Honorio Salmeron Valdivieso}
Graduated in Industrial Engineering Technologies by the Malaga's University. Partial studies of Physical Sciences on Granada's University.\\ \href{mailto:honosalval@gmail.com}{\textit{honosalval@gmail.com}}
\end{IEEEbiography}



\begin{thebibliography}{10}
\bibitem{Walton}
D S Meek and D J Walton, \emph{Approximation of discrete data by $G^1$ arc splines}. \hskip 1em plus 0.5em minus 0.4em\relax Computer-Aided Design Volume 24, Issue 6, June 1992, Pages 301-306.
\bibitem{Alla}
Alla Safonova, Jarek Rossignac,\emph{Compressed piecewise-circular approximations of 3D curves},    Computer-Aided Design,    Volume 35, Issue 6,    2003,    Pages 533-547,
\bibitem{Milan}
Milan K Yeung and Desmond J Walton, \emph{Curve fitting with arc splines for NC toolpath generation}. \hskip 1em plus  0.5em minus 0.4em\relax Computer-Aided Design Volume 26, Number 11, November 1994, Pages 845-849.
\bibitem{Alex4}
A. Gribov, "Searching for a Compressed Polyline with a Minimum Number of Vertices," 2017 14th IAPR International Conference on Document Analysis and Recognition (ICDAR), Kyoto, 2017, pp. 13-14.
doi: 10.1109/ICDAR.2017.254
\bibitem{Ji}
Ji-Hwei Horng and Johnny T. Li, \emph{A dynamic programming approach for fitting digital planar curves with line segments and circular arcs}. \hskip 1em plus  0.5em minus 0.4em\relax Pattern Recognition Letters 22 (2001) 183-197.

\bibitem{Liu}
Liu Yin, Yu Yajie and Liu Wenyin, "Online segmentation of freehand stroke by dynamic programming," Eighth International Conference on Document Analysis and Recognition (ICDAR'05), Seoul, South Korea, 2005, pp. 197-201 Vol. 1.
\bibitem{Xi}
Xinghua Song, Martin Aigner, Falai Chen, Bert Jüttler \emph{Circular spline fitting using an evolution process}. \hskip 1em plus  0.5em minus 0.4em\relax Journal of Computational and Applied Mathematics 231 (2009) 423-433

\bibitem{JGD}
JAMES GEORGE DUNHAM, \emph{Optimum uniform piecewise linear approximation    of planar curves}. \hskip 1em plus  0.5em minus 0.4em\relax Journal of Computational and Applied Mathematics 231 (2009) 423-433

\bibitem{Alex}
Alexander Gribov, \emph{Optimal Compression of a Polyline with Segments and Arcs}. \hskip 1em plus  0.5em minus 0.4em\relax arXiv:1604.07476.

\bibitem{Alex2}
Alexander Gribov, \emph{Approximate Fitting of Circular Arcs when Two Points are Known}. \hskip 1em plus  0.5em minus 0.4em\relax arXiv:1504.06582.

\bibitem{Alex3}
Eugene Bodansky and Alexander Gribov, \emph{Approximation of a Polyline with a Sequence of Geometric Primitives}. \hskip 1em plus  0.5em minus 0.4em\relax International Conference Image Analysis and Recognition. ICIAR 2006: Image Analysis and Recognition pp 468-478.

\bibitem{Taubin}
G. Taubin, \emph{Estimation of planar curves, surfaces, and nonplanar space curves defined by implicit equations with applications to edge and range image segmentation}, in IEEE Transactions on Pattern Analysis and Machine Intelligence, vol. 13, no. 11, pp. 1115-1138, Nov 1991.

\bibitem{Kasa}
I. Kasa, \emph{A curve fitting procedure and its error analysis}, IEEE Trans. Inst. Meas., Vol. 25, pages 8-14, (1976)
\end{thebibliography}
\end{document}